\documentclass[aps,pra,reprint,amsmath,amssymb,twocolumn,showpacs,superscriptaddress]{revtex4}
\usepackage{bbm}
\usepackage{graphicx}
\usepackage{amsmath}
\usepackage{amssymb,amsthm}
\usepackage{color}
\usepackage[sans]{dsfont}
\usepackage{hyperref}

\newcommand{\ket}[1]{|#1\rangle}

\newcommand{\bra}[1]{\langle#1|}

\newcommand{\refchiN}{6}
\newcommand{\refnewton}{7}

\def\eq{\begin{eqnarray}}
\def\en{\end{eqnarray}}

\def\beq{\begin{eqnarray}}
\def\een{\end{eqnarray}}
\def\bfig{\begin{figure}}
\def\efig{\end{figure}}

\begin{document}

\title{Two-boson composites}

\author{Malte C. Tichy}
\affiliation{Department of Physics and Astronomy, University of Aarhus, DK-8000 Aarhus C, Denmark}

\author{P. Alexander Bouvrie}
\affiliation{Departamento de F\'isica At\'omica, Molecular y Nuclear and Instituto Carlos I de F\'isica Te\'orica y Computacional, Universidad de Granada, E-18071 Granada, Spain}

\author{Klaus M\o{}lmer}
\affiliation{Department of Physics and Astronomy, University of Aarhus, DK-8000 Aarhus C, Denmark}


\date{\today}

\begin{abstract}
Composite bosons made of two bosonic constituents exhibit deviations from ideal bosonic behavior due to their substructure. This deviation is reflected by the \emph{normalization ratio} of the quantum state of $N$ composites. We find a set of saturable, efficiently evaluable bounds for this  indicator,  which quantifies the bosonic behavior of composites via the entanglement of their constituents. We predict  an abrupt  transition between ordinary and exaggerated bosonic behavior in a condensate of two-boson composites.
\end{abstract}
\date{\today}
\maketitle

\emph{Introduction.} The (fermionic) bosonic behavior of any elementary or composite particle  is ultimately implied by the spin-statistics theorem \cite{fierz1939,Pauli1940}, which can be derived under many different assumptions \cite{Jabs2009}. For composite bosons made of two fermions, the Pauli principle that acts on the constituents  {modifies the ideally expected bunching behavior } \cite{avancini,Rombouts2002,Sancho2006}, and changes the bosonic commutation relation \cite{Law2005}. The statistics of composites was recently re-considered from the perspective of quantum information \cite{Law2005}.  Both, in the many-body properties of Bose-Einstein-Condensates (BECs)   \cite{Combescot2003,Chudzicki2010,Ramanathan2011,Combescot2011,Combescot2008a,Combescot2010,Tichy2012CB,Gavrilik2012,Gavrilik2013} and in dynamical processes \cite{Brougham2010,Tichy2012a,Ramanathan2011a,Thilagam2013}, \emph{entanglement} between two fermionic constituents turns out to be  the crucial ingredient to ensure bosonic behavior \cite{Law2005}. 

 While for atoms and molecules, the impact of the Pauli principle that acts on the constituent electrons  is typically small \cite{Chudzicki2010}, the question of the effective compositional hierarchy and the impact of bosonic and fermionic effects on a higher level remains open, e.g.,~for molecules made of two bosonic atoms, and it is lively debated for $\alpha$-particles in nuclear physics \cite{Zinner2008,Funaki2010,Zinner2013}. For a composite boson made of two bound \emph{bosonic} constituents,  no Pauli-blocking jeopardizes the multiple occupation of single-particle states. One could therefore expect such compound to simply \emph{inherit} the bosonic nature of its own constituents.   However, as we show below, the behavior of two-boson composites can heavily deviate from the ideal, because the single-particle states of the constituents  tend to be unusually often multiply populated, leading to a \emph{super}-bosonic compound. Although all matter is ultimately made of fermions, any high-level composite that is made of two bosonic constituents will face such super-bosonic effects.

The quantitative indicator for bosonic features in the many-body theory of composites is the \emph{composite-boson normalization ratio} $\chi_{N+1}/\chi_N$ \cite{Law2005,Combescot2011,Combescot2003,Chudzicki2010,Ramanathan2011}.  However, even when the two-boson wavefunction is  known, the  complexity of the algebraic expression for $\chi_{N+1}/\chi_N$ renders an evaluation for large $N$ unfeasible \cite{Law2005}.

Here, we solve this  problem by providing tight, saturable bounds for the normalization ratio, which allow us to efficiently characterize two-boson composites via three easily accessible quantities: the number of composites  $N$, and the purity $P$ and the largest eigenvalue $\lambda_1$ of the reduced density matrix of one constituent boson,  which can be obtained from the two-boson wavefunction. This allows a  quantitative discussion of the bosonic behavior of two-boson composites in terms of entanglement measures: The geometric measure of entanglement is connected to $\lambda_1$ via $E_G=1-\lambda_1$ \cite{PRA68Godbart}, the Schmidt number fulfills $K=1/P$ \cite{Horodecki2009}. In contrast to two-fermion composites,  biboson composites exhibit exaggerated bunching. As a remarkable consequence, an abrupt transition takes place between ordinary bosonic behavior and a super-condensation regime in which the extraordinary bunching tendency of the constituents dominates and the condensation of the constituent parts competes with the condensation of the  composite whole.

\begin{figure}[ht]
\includegraphics[width=.9\linewidth]{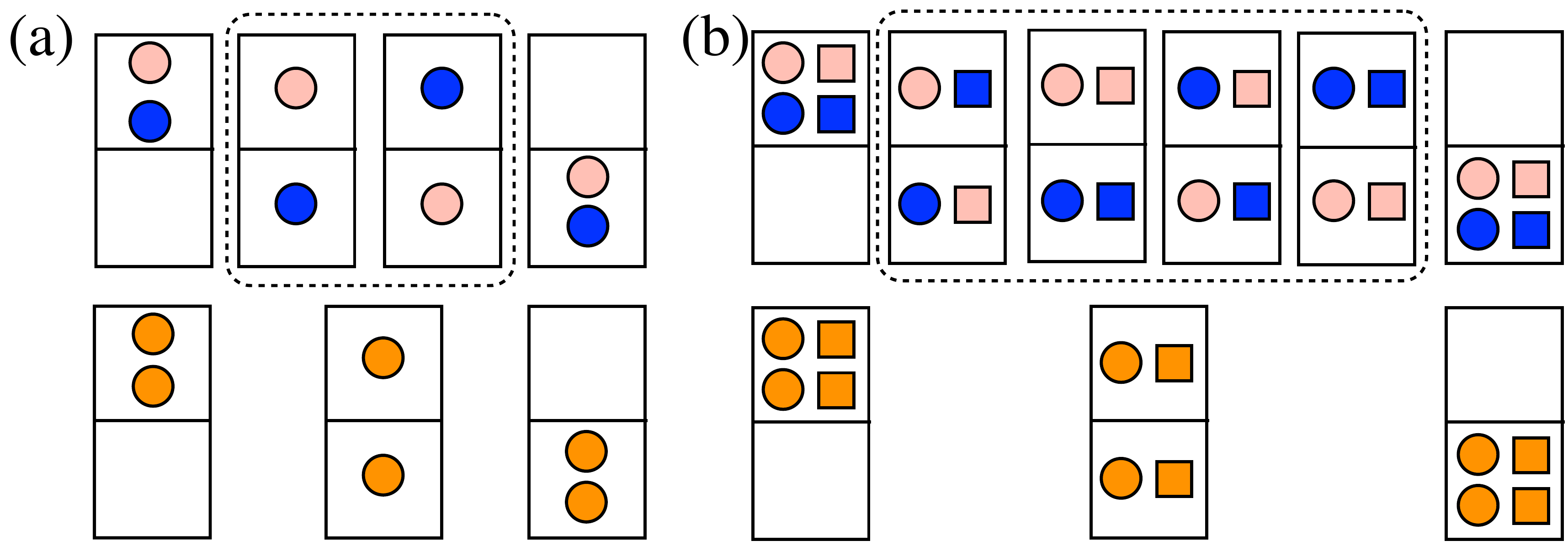}
\caption{(color online) (a) Two distinguishable particles in two states lead to four distinguishable states. Making the particles indistinguishable merges the two states with one particle per state. Therefore, Bose-Einstein statistics favors states with two bosons in one state with respect to the statistics of distinguishable particles.  (b) For inseparable strongly bound  bibosons (made of a circle and a square), the bunching tendency is enhanced, since \emph{four} a priori distinguishable states are merged when the particles become indistinguishable. }
\label{FigCounting.pdf}
\end{figure}

\emph{Biboson bosons.}
Every composite made of two distinguishable elementary bosons can be described by a wavefunction $\Psi(\vec r_a, \vec r_b)= \sum_{j=1}^S \sqrt{\lambda_j} \phi_j(\vec r_a) \psi_j(\vec r_b)$, expanded on Schmidt mode functions $\phi_j, \psi_j$ \cite{Horodecki2009,Law2005}. In second quantization, the creation of a composite boson is described by
\eq
\hat c^\dagger=\sum_{j=1}^S \sqrt{\lambda_j} ~ \hat a^\dagger_j \hat b^\dagger_j ,~~  \label{eq:ddefin} \lambda_1 \ge \lambda_2 \ge \dots  \ge 0, ~~ \sum_{j=1}^S \lambda_j =1,
\en
where $\hat a^\dagger_j$ ($\hat b^\dagger_j$) creates a boson in $\phi_j (\psi_j)$.
The creation of a biboson is described by $\hat d^\dagger_j :=  \hat a^\dagger_j \hat b^\dagger_j$,  which fulfils
\eq \left[  \hat d_j , \hat d_k  \right] & =&\left[  \hat d_j^\dagger , \hat d^{\dagger}_k  \right]  =0, \\
 \left[  \hat d_j , \hat d^{\dagger}_k  \right] &=& \delta_{j,k} (1+2 ~ \hat n_{j} ) , \label{commutationDibosons} \\
 \left( \hat d^\dagger_j \right)^N \ket{0} &=&  N! ~ \ket{N}_j ,  \label{overnormalization}
\en
where $\hat n_j$ counts the number of bibosons in the $j$th mode. Bibosons tend to bunch more strongly than bosons, which is reflected by (\ref{commutationDibosons}) and by the  over-normalization of the $N$-biboson state (\ref{overnormalization}). This enhanced bunching tendency is ultimately rooted in the larger number of states that coincide under symmetrization of a state of distinguishable particles, as illustrated in Fig.~\ref{FigCounting.pdf}.

An $N$-composite state is obtained by the $N$-fold application of the creation operator (\ref{eq:ddefin}) on the vacuum,
\eq
\ket{N} =\frac{  \left( \hat c^\dagger \right)^N }{ \sqrt{\chi_N~ N!} }\ket{0}  , \label{NCompositeState}
\en
where $\chi_N \ge 1$ is the composite boson \emph{normalization factor} \cite{Law2005}, which accounts for the over-normalization of those components of the wavefunction for which some Schmidt modes are occupied by more than one biboson. This factor is the complete homogeneous symmetric polynomial of degree $N$ in the Schmidt coefficients $\vec \Lambda=(\lambda_1, \dots , \lambda_S)$ \cite{MacDonaldSymmetric}:
\eq
\chi_N&=& N! \sum_{1 \le p_1 \le \dots \le p_N \le S} \prod_{k=1}^N \lambda_{p_k}
 \label{chifdef} ,
\en
where terms with $p_{n}=\dots =p_{n+m}$ allow for multiply occupied modes.
A variant of the Newton-Girard identity for symmetric polynomials \cite{Ramanathan2011,MacDonaldSymmetric} leads to the recursion
\eq
 \chi_N &=&(N-1)! \sum_{m=1}^{N} \frac{\chi_{N-m}}{(N-m)!} M(m) , \label{RecursionBosons}
\en
where we introduced the $m$th power-sum
\eq
 M(m) &=& \sum_{j=1}^S \lambda_j^m . \en

The \emph{normalization ratio} $\chi_{N+1}/\chi_N$ \cite{Law2005} determines the bosonic quality of a state of $N$  biboson composites, e.g., for  the expectation value of the commutator \cite{Law2005,Ramanathan2011,Combescot2011},
\eq
\bra N  \left[ \hat c, \hat c^\dagger \right] \ket N = 1+  2\sum_{j=1}^S \lambda_j   \bra N \hat n_j  \ket N =2 \frac{\chi_{N+1}}{\chi_N} -1 ,
 \en
which implies that ideal bosons fulfil $\chi_{N+1}/\chi_N =1$. While the normalization ratio for bifermion bosons  decreases monotonically with $N$ \cite{Chudzicki2010}, we find the opposite behavior for biboson composites:  \eq
1 \stackrel{(a)}{\le}  \frac{\chi_N}{\chi_{N-1}} \stackrel{(b)}{\le}  \frac{\chi_{N+1}}{\chi_{N}} \stackrel{(c)}{\le} \frac{N+1}{N} \frac{\chi_{N}}{\chi_{N-1}} \stackrel{(d)}{\le} N+1 , \label{hierarchy1}
 \en
 where $(a)$ and $(d)$ are implied by the normalization of Schmidt coefficients (\ref{eq:ddefin}), and proofs for $(b)$ and $(c)$ can be found in  Refs.~\cite{Hardy1988} and \cite{BanksMartin2013}, respectively. In terms of the occupation of the Schmidt modes, the $N$-composite state (\ref{NCompositeState}) reads
\eq
\ket{N}  =  \sqrt{\frac{ N!}{ \chi_N} } \sum_{m_1, \dots, m_S \ge 0}^{\sum_{j=1}^S m_j = N} \left[ \prod_{j=1}^S \sqrt{\lambda_j}^{m_j}  \right] \ket{m_1 \dots m_S}  , \label{counting}
\en
where the unusual statistics of the bibosons becomes apparent through the absence of the normal combinatorial factors, which are compensated by the extraordinary bunching-pre-factors in (\ref{overnormalization}).

\emph{Bounds on the normalization factor.}
The  evaluation of $\chi_N$ through (\ref{chifdef}) or (\ref{RecursionBosons}) scales prohibitively with $N$, even when shortcuts due to multiplicities of Schmidt coefficients are exploited \cite{Supplementary}. To permit a quantitative discussion of two-boson composites, reliable bounds and approximations to $\chi_N$ are necessary.

 For bifermion composites, the purity $P\equiv M(2)$ of the reduced density matrix of one constituent fermion  turns out to be an excellent indicator for bosonic behavior as long as $N \ll 1/P$ \cite{Chudzicki2010,Ramanathan2011,Tichy2012CB}, since the influence of Pauli-blocking is  largely governed by $P$. For biboson composites, however, a Schmidt mode can be multiply occupied, which induces exaggerated bunching in that mode, driven by the occupation-dependent commutator (\ref{commutationDibosons}). The most populated Schmidt mode will eventually dominate the composites, and we  expect the normalization factor of two distributions with the same purity $P$ but different largest Schmidt coefficient $\lambda_1$ to differ dramatically.

\begin{figure}[ht]
\includegraphics[width=1\linewidth]{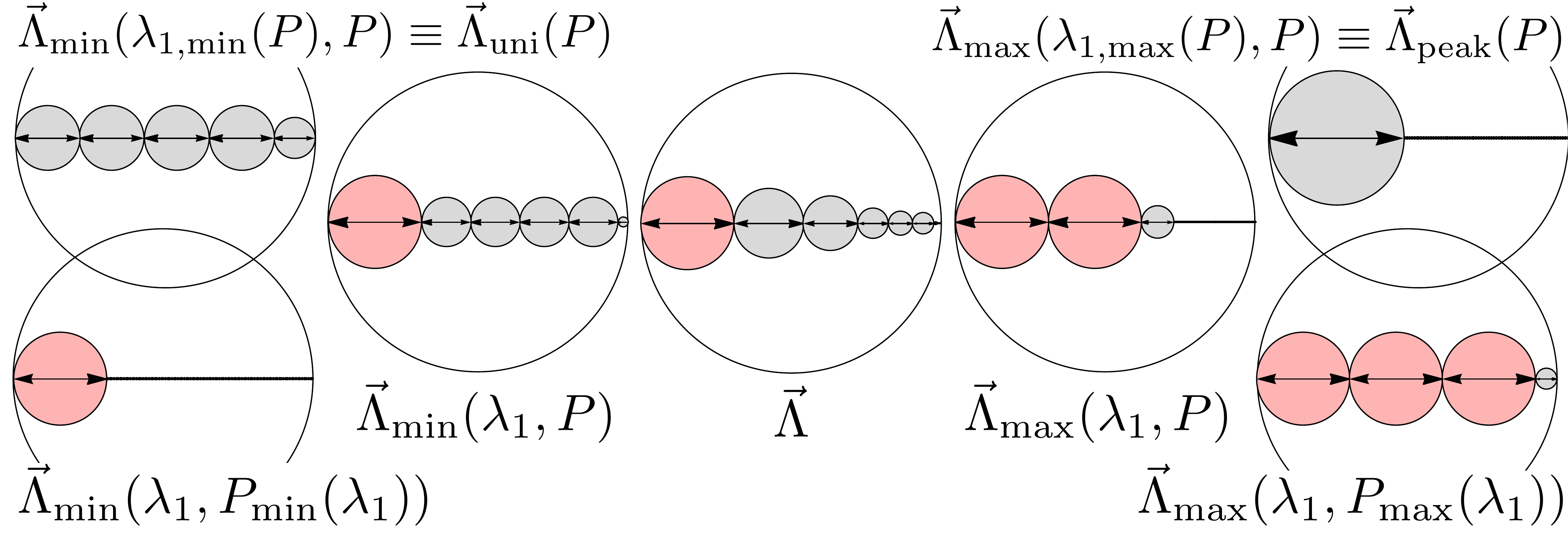}
\caption{(color online) Original and  modified distributions that limit the normalization factor. The diameters of the circles correspond to the magnitude of the respective Schmidt coefficient, such that the fraction of filled area represents the purity $P$ of the respective distribution. A distribution $\vec \Lambda$, with $\lambda_1=0.31$ and $P=0.205$ (center) leads to a $\chi_N$ that is bound by the $\chi_N$ belonging to  distributions with large multiplicities of the Schmidt coefficient magnitudes. All circles that symbolize a coefficients $\lambda_1$ are drawn in red, the left-right ordering reflects the hierarchy of the bounds expressed by Eqs.~(\ref{superbound},\ref{Pbounds},\ref{lam1bounds}). }
\label{FigureDistributions.pdf}
\end{figure}
A remedy is our following bound for $\chi_N$  in   $\lambda_1$ \emph{and}   $P$:   From Eq.~(\ref{RecursionBosons}), we see that $\chi_N$  is monotonically increasing in  all   power-sums $M(m)$, since all appearing pre-factors
 are non-negative. Therefore, those distributions $\vec \Lambda_{\text{max}(\text{min})}$ with largest Schmidt coefficient $\lambda_1$ and purity $P$ that maximize (minimize) power-sums $M(m)$ also maximize (minimize) $\chi_N$ and $\chi_{N+1}/\chi_N$ \cite{Supplementary}. We construct these distributions (see Fig.~\ref{FigureDistributions.pdf}) and determine their corresponding $\chi_N$ explicitly: By virtue of (\ref{eq:ddefin}),  the unknown Schmidt coefficients $\lambda_{j\ge 2}$ fulfil
\eq
\sum_{j=2}^S \lambda_j = 1-\lambda_1; ~ \sum_{j=2}^S \lambda_j^2 = P-\lambda_1^2; ~ 0 \le \lambda_j \le \lambda_1 \label{constraintsonlam} .
\en
Under this constraint,  higher-order power-sums ${M(m\ge 3)}$ are maximized by $\vec \Lambda_{\text{max}}( \lambda_1, P)$, with  \cite{Supplementary}
\eq \lambda_1=\lambda_2=\dots = \lambda_{L-1} \ge  \lambda_L \ge \lambda_{L+1} = \dots \lambda_S ,  \label{maxordering} \en
and $L=\left\lceil P/\lambda_1^2 \right\rceil$, where we assume  the limit $S \rightarrow \infty$, such that only the first $L$ coefficients remain finite, while all others converge to zero. Conversely, power-sums are minimized by
 $\vec \Lambda_{\text{min}}(\lambda_1,P )$,  with \cite{Supplementary}
  \eq \lambda_1 \ge \lambda_2=\dots = \lambda_{S-1} \ge \lambda_S ,~S=1 + \left \lceil \frac{(1-\lambda_1)^2}{P-\lambda_1^2} \right \rceil . \label{minordering}   \en
The normalization factor $\chi_N$ of any distribution $\vec \Lambda$ fulfils
\eq
\chi_N^{\vec \Lambda_{\text{min}}(\lambda_{1},P) } \le   \chi_N \le
\chi_N^{\vec \Lambda_{\text{max}}(\lambda_{1},P)} .  \label{superbound}
\en
An analogous hierarchy applies also for the normalization \emph{ratio} $\chi_{N+1}/\chi_{N}$. Since  $\vec \Lambda_{\text{min(max)}}(\lambda_1,P)$
 contain at most three different non-vanishing Schmidt coefficients $\lambda_j$,
the bounds in (\ref{superbound}) can be evaluated easily as sums over incomplete $\Gamma$-functions  \cite{Olver:2010:NHMF,Supplementary}.

\begin{figure}[t]
\includegraphics[width=1\linewidth]{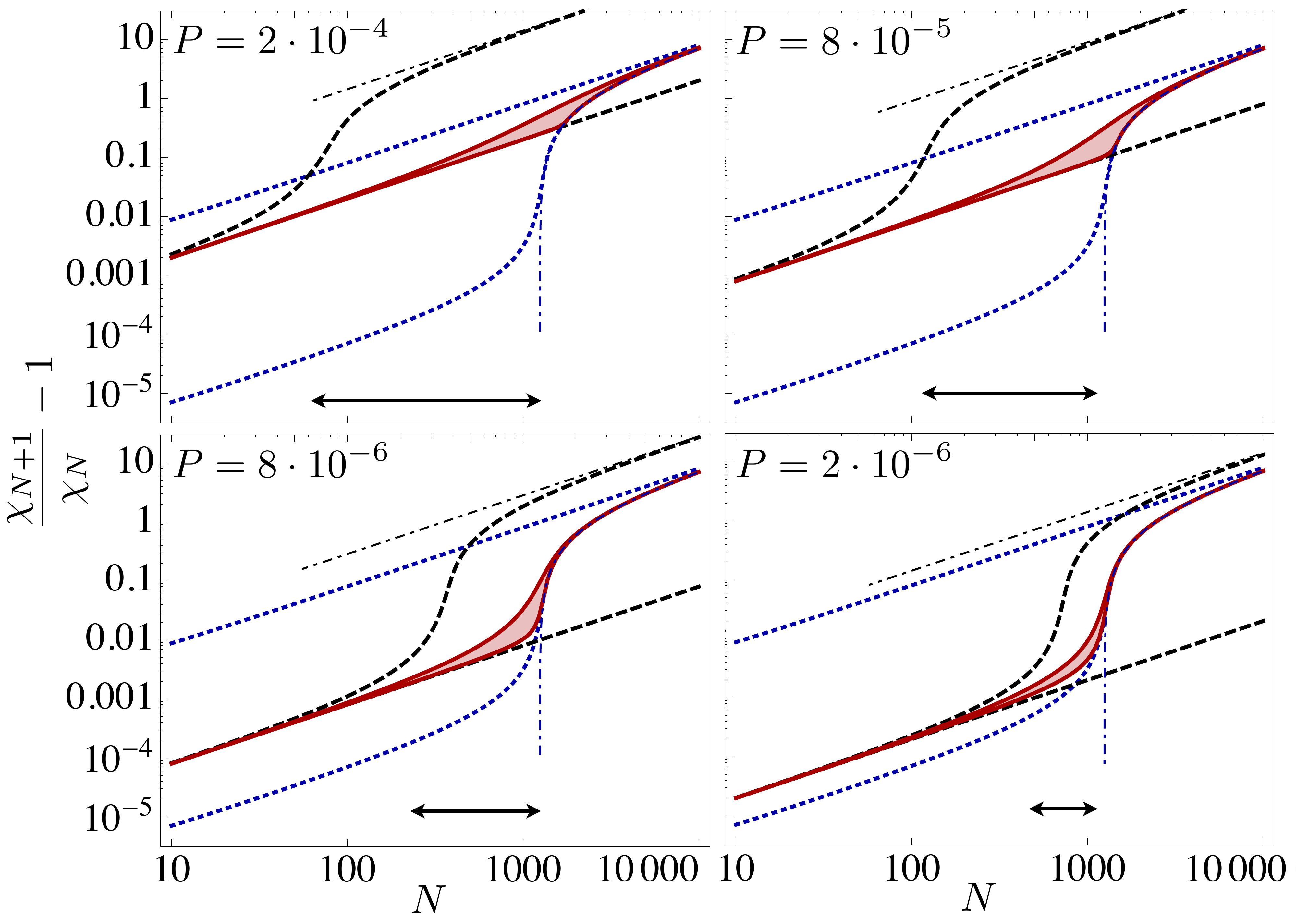}
\caption{(color online) Deviation from ideal normalization ratio, $\chi_{N+1}/\chi_{N}-1$, as a function of $N$, for $\lambda_1=8 \cdot 10^{-4}$ and different values of $P$. The tight bounds given by (\ref{superbound}) are shown as a red solid line which encloses a shaded area; the $P$-dependent bounds (\ref{Pbounds}a) are represented as a black dashed line, the $\lambda_1$-dependent bounds (\ref{lam1bounds}d) are given as blue dotted lines (identical in each subplot). The thin dash-dotted lines show the approximations (\ref{Pbounds}b) and  (\ref{lam1bounds}c), which become efficient only for large $N$. The arrows indicate the range of $N$ for which the weak bounds are inefficient: $1/\sqrt{P} \le N \le 1/\lambda_1 $.
}
\label{NewFigBounds.pdf}
\end{figure}
We can infer weaker, however very instructive, bounds that depend uniquely on $\lambda_1$ \emph{or} $P$. For this purpose, we find the possible intervals of
 $\lambda_1$ and $P$,
\eq
P \le \lambda_{1,\text{min}}(P) & \le&  \lambda_1  \le \lambda_{1,\text{max}}(P) =  \sqrt{P} , \\
 \lambda_1^2 = P_{\text{min}}(\lambda_1) &\le&  P \le P_{\text{max}}(\lambda_1) \le \lambda_1,
\en
where
\eq
\lambda_{1,\text{min}}(P) &=& \frac{1}{S} \left( \sqrt{\frac{P S-1}{{S-1}}}+1 \right), ~~
S=\left\lceil\frac 1 P \right\rceil  \nonumber , \\
 P_{\text{max}}(\lambda_1) &=&\lambda_1^2 \left\lfloor \frac 1 {\lambda_1} \right\rfloor + \left( 1- \lambda_1 \left\lfloor \frac 1 {\lambda_1} \right\rfloor  \right)^2 \label{PAndLambdaMaxMin} .
\en
Lower (upper) bounds in $P$ -- independent of $\lambda_1$ -- are obtained for  $\lambda_1 = \lambda_{1,\text{min}(\text{max})}(P)$, for which the maximizing and minimizing distributions coincide with the \emph{uniform} (\emph{peaked}) distribution \cite{Tichy2012CB}, $\vec \Lambda_{\text{max}}=\vec \Lambda_{\text{min}}=\vec \Lambda_{\text{uni}(\text{peak})}$, which contain at most two distinct Schmidt coefficients $\lambda_j$.  For these simple distributions,  we find  \cite{Supplementary},
\eq
P N+1  \stackrel{(a)}{\le} \frac{\chi_{N+1}}{\chi_N}  \stackrel{(a)}{\le}  \sqrt{P}  \frac{   \Gamma ( N+2 ,  \frac {1-\sqrt P} {\sqrt{P}}  ) }{ \Gamma (  N+1 , \frac {1-\sqrt P } {\sqrt{P}} ) }   \label{Pbounds}
\stackrel{(b)}{\le} \sqrt P N +1 ,   \hspace{6pt}
\en
where $\Gamma \left( s, x \right) $ is the incomplete $\Gamma$-function \cite{Olver:2010:NHMF}, which saturates $(b)$ for $N \gg 1/\sqrt{P}$ \cite{Supplementary}.

On the other hand, lower (upper) bounds in $\lambda_1$ -- independent of $P$ -- are obtained for $P = P_{\text{min(max)}}(\lambda_1)$,
\eq
\lambda_1 (N+1) \stackrel{(c)}{\le }    \lambda_1 \frac{\Gamma (N+2,\frac{1-\lambda_1}{\lambda_1})}{\Gamma (N+1,\frac{1-\lambda_1}{\lambda_1})} \stackrel{(d)}{\le}  \frac{\chi_{N+1}}{\chi_N} \stackrel{(d)}{\le}  \lambda_1 N+1 , \hspace{6pt}  \label{lam1bounds}
\en
where $(c)$ becomes efficient for $N \gg 1/\lambda_1$. The bounds are shown in Fig.~\ref{NewFigBounds.pdf}, where we can read off three different regimes: For $N < 1/\sqrt{P}$, the bounds in $P$ (black dashed, Eq.~(\ref{Pbounds}a)) are efficient and $\chi_{N+1}/\chi_{N}-1<1$, i.e.~the composites behave rather bosonically. For $N>1/\lambda_1$, we have $\chi_{N+1}/\chi_{N}-1>1$, the composites are super-bosonic, and the largest Schmidt coefficient dominates the picture, making the bounds in $\lambda_1$ (blue dotted, Eq.~(\ref{lam1bounds}d)) efficient. In the intermediate region, $1/\sqrt{P} < N < 1/\lambda_1$, both simple bounds are inefficient. When we keep the dependence on, both, $P$ \emph{and} $\lambda_1$,  Eq.~(\ref{superbound}) gives a significantly tighter  interval (red solid). Towards smaller values of $P$ and $\lambda_1$ (and larger pertinent composite particle numbers $N$), the bounds in (\ref{Pbounds}) and (\ref{lam1bounds}) become more and more step-like and the transition between the regimes more abrupt.

\begin{figure}[t]
\includegraphics[width=.9\linewidth]{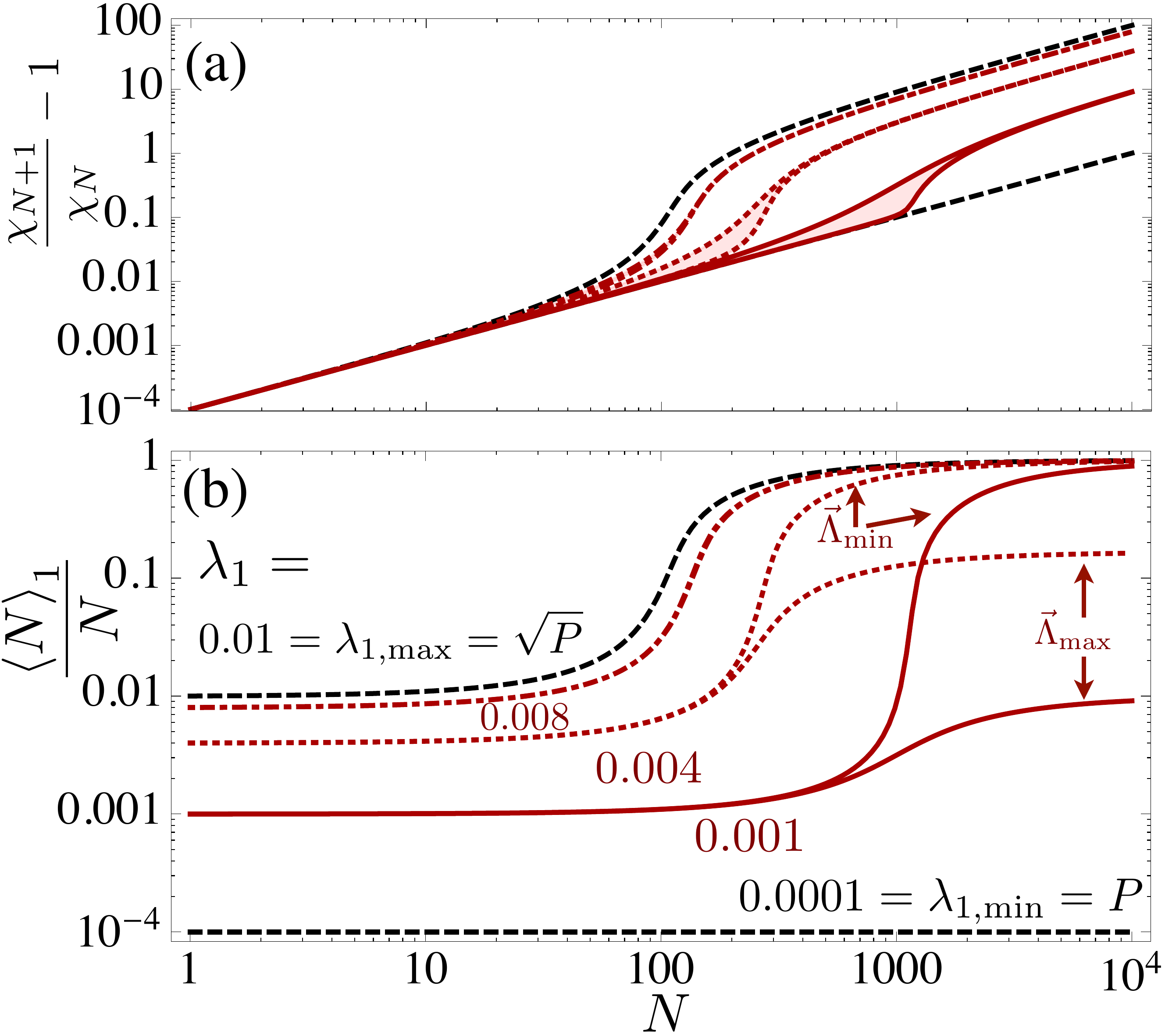}
\caption{(color online) (a) Upper and lower bounds to ${\chi_{N+1}/\chi_{N}-1}$,  for $P=10^{-4}$ and different values of $\lambda_1$ (as given in the lower panel, for $\lambda_1=\lambda_{1,\text{max}}$ and $\lambda_1=\lambda_{1,\text{min}}$ the bounds coincide). (b) Average fraction of particles $\langle N \rangle_1/N$ in one Schmidt mode with magnitude $\lambda_1$, for the maximizing (minimizing) distribution $\vec \Lambda_{\text{max(min)}}(\lambda_1,P)$. For $N \ll 1/\lambda_1$, the fraction amounts to $\lambda_1$. At $\langle N \rangle_1 \approx 1$, the transition to super-condensation takes place: More and more bibosons populate a Schmidt mode with magnitude $\lambda_1$, eventually saturating at unity for the minimizing distribution (for which one Schmidt mode fulfils $\lambda_j=\lambda_1$), and at $1/(L-1)$ for the maximizing distribution (since $L-1$ Schmidt coefficients adopt the value $\lambda_1$, see (\ref{maxordering})). For the five different values of $\lambda_1$, the respective multiplicities of $\lambda_1$ in the maximizing distribution $\vec \Lambda_{\text{max}}(\lambda_1, P)$ are $1,1,6,100,10000$.}
\label{ChiNandAvPN.pdf}
\end{figure}

\emph{Counting statistics.}
The occupation of the most prominent Schmidt mode explains the transition between the two regimes.   For non-interacting bosons and distinguishable particles that are distributed among $S$ modes, the average number of particles in the first mode is  $\langle N \rangle_1 =N \lambda_1$. For bibosons, however, this relation is no longer true, and the \emph{average} population of each Schmidt mode depends on the \emph{total} number of particles in the system: Although non-\emph{interacting}, due to the population-dependent commutator (\ref{commutationDibosons}), bibosons are \emph{not independent}! From Eq.~(\ref{counting}), we infer the probability to find $m_1$ bibosons in the first Schmidt mode,
\eq
P(m_1) &=&  \lambda_1^{m_1}  \frac{N!}{(N-m_1)!} \frac{ \chi_{N-m_1}^{ [ \lambda_2 \dots \lambda_S ] }  }{\chi_N^{ [ \lambda_1, \lambda_2 \dots \lambda_S ] }} , \label{countinstat}
\en
and obtain the average number of particles in that mode, $\langle N \rangle_1 = \sum_{m_1=0}^{N} m_1 P(m_1) .$ The fraction of particles in the first Schmidt mode $\langle N \rangle_1/N$ is shown as a function of the total particle number $N$ in Fig.~\ref{ChiNandAvPN.pdf} (b). The occupation jumps abruptly at $N \approx 1/\lambda_1$ from the initial combinatorial value $\lambda_1$ to $1/(L-1)$, where $(L-1)$ is the multiplicity of $\lambda_1$ in the respective distribution.  The jump takes place precisely at the value of $N$ at which also the normalization ratio $\chi_{N+1}/\chi_N$ jumps to larger values (Fig.~\ref{ChiNandAvPN.pdf} (a)).

When $N \ll 1/\lambda_1$, the population of the first Schmidt mode is small, and $\langle N \rangle_1 / N \approx \lambda_1$, just as for bosons and distinguishable particles -- the probability for two particles to populate the same mode is negligible, and the bunching of bibosons can be neglected. Although the composites condense in the state $\Psi$, their constituents do not populate any state macroscopically.  When we increase the number of composites $N$ (or, alternatively, when we increase $\lambda_1$), the first Schmidt mode is populated with a non-vanishing number of bibosons as soon as $N \gtrsim 1/\lambda_1$. Then, a ``winner-takes-it-all''-effect takes place: Due to the occupation-dependent commutator (\ref{commutationDibosons}), an already occupied mode is  likely to be populated further.  Therefore, the first mode to attain a sizeable population (i.e.~mode 1) attracts all subsequently added particles. Eventually, the overwhelming majority of bibosons populate the first Schmidt mode, which induces the sudden change in the normalization ratio.
Consistently, no combinatorial factors are present in (\ref{counting}), which effectively privileges the population of the first Schmidt mode. When all Schmidt coefficients $\lambda_k$ are identical, as for the uniform distribution, no mode can be privileged with respect to the others, due to symmetry. Therefore, the occupation of the first Schmidt mode does not increase with $N$ for $\lambda_1=\lambda_{1,\text{min}}=P$ in Fig.~\ref{ChiNandAvPN.pdf}, whereas the counting statistics (\ref{countinstat}) still differs from the binomial statistics of distinguishable particles.

\emph{Conclusions and outlook. } In Eq.~(\ref{superbound}), we provide an easily evaluable bound for the normalization factor and normalization ratio of biboson composites, which can be readily applied to any composite system to clarify whether the composite boson under consideration can be treated as ideal. By deliberately leaving the realm of ideal bosons, we observe a transition between two well-defined regimes:  When $N \sqrt{P} \ll 1$, we have $\chi_N \approx 1$, and a condensate of biboson composites can be treated as a BEC of ordinary bosons with negligible substructure. For  $N \lambda_1 \gg 1$, however, Schmidt modes with magnitude $\lambda_1$ are macroscopically populated, the resulting \emph{super-condensate} is governed by the super-bunching tendency of bibosons. In the super-condensate, not only do all composites condense in the same state $\Psi$, but on a \emph{subordinate} level, also their constituents condense! To observe this BEC-BEC$^2$ transition, the macroscopic population of the Schmidt modes with the highest expansion coefficients may be addressed by a suitable species-selective probe.

We expect super-bunching of bibosons (see Fig.~\ref{FigCounting.pdf}) to facilitate the BEC of composites, which may occur at lower densities and higher temperatures than for a gas of elementary bosons. Remarkably, this provides an original means to probe the two-particle wave-function of the constituents non-destructively, via the statistical behavior of the composite \cite{Tichy2012a}.
 Composites made of two identical bosons, or of three or even more bosonic constituents will behave in an even more violently super-bosonic way: The resulting commutator of multi-boson operators (\ref{commutationDibosons}) and the over-normalization (\ref{overnormalization}) will be enhanced further. The absence of a general Schmidt decomposition for three-particle states \cite{Horodecki2009}, however, renders an investigation beyond the two-constituent paradigm difficult. An approach via entanglement measures for multipartite entanglement seems advisable \cite{Horodecki2009}, and could further tighten the connection between composite-particle physics and quantum information. Another desideratum is the interference \cite{Thilagam2013,Tichy2012a,Ramanathan2011a,Brougham2010,Lee:2013fk} and the statistical behavior \cite{Combescot,Kaszlikowski2013,Thilagam2013b} of biboson composites, which can be approached via the bounds presented here. The occupation-dependent commutator (\ref{commutationDibosons}), however, makes a  formal treatment difficult.  

Physical composition is a hierarchical property, and we expect a subtle interplay between the Pauli principle that acts on all elementary fermions and the super-bunching induced by constituent bosons: For example, two fermions may be combined to form a composite boson, two such bosons may then be joined to another superordinate compound. Depending on the resulting four-fermion-state, the emerging composite needs to be treated as a ``perfect'' boson, as a super-bosonic two-boson compound, or as a sub-bosonic four-fermion aggregate. The  normalization ratio then indicates which description is more appropriate, and may contribute, e.g., to the debate on $\alpha$-particle condensation \cite{Zinner2008,Funaki2010,Zinner2013}.

\emph{Acknowledgements.}  The authors would like to thank Florian Mintert, Alagu Thilagam and Nikolaj Thomas Zinner for stimulating discussions and for valuable comments on the manuscript.  M.C.T. gratefully acknowledges support by the Alexander von Humboldt-Foundation through a Feodor Lynen Fellowship. K.M. gratefully acknowledges support by the Villum Foundation. P.A.B. gratefully acknowledges support by the Progama de Movilidad Internacional CEI BioTic en el marco PAP-Erasmus, the MINECO grant FIS2011-24540 and the excellence grant FQM-7276 of the Junta de Andaluc\'ia.


\begin{widetext}

\section*{Supplemental material}

\section{Algebraic properties of the normalization factor}
In order to make this Supplemental Material self-contained, we summarize useful algebraic relations for the normalization factor $\chi_N$, defined as Eq.~(\refchiN) in the main text. We define 
\eq 
 \Xi \{ x_1, \dots , x_N \}   = N! \sum_{1 \le p_1 \le \dots \le p_N \le S} \prod_{k=1}^N \lambda_{p_k}^{x_k}  \label{xidef}
\en
such that
\eq 
\chi_N =  \Xi \{ \underbrace{1 \dots 1}_N  \} . 
\en
This representation of $\chi_N$ allows us to formulate a useful recursive relation, 
\eq 
\label{RecursionBosons2}
\Xi \{ x, \underbrace{1 \dots 1}_{K} \}  
 = M(x) ~{ \Xi}\{ \underbrace{  1 \dots 1 }_{K} \}  +  K~{ \Xi}\{ x+1, \underbrace{1\dots 1 }_{K-1} \} .
\en
Possible multiplicities of Schmidt coefficients are beneficial for evaluation, since these can be exploited via 
\eq
\chi_N^{\vec \Lambda}   = \lambda^{N} \frac{(N+S-1)!}{(S-1)!} \label{singlelam2} ~ \text{ for } \vec \Lambda= (\underbrace{ \lambda  \dots \lambda }_{S}) ,
\en
and via 
\eq 
\chi_N^{[ \lambda_1 \dots \lambda_S ] }  =  \sum_{M=0}^N \chi_M^{[ \lambda_1  \dots \lambda_L ]} ~ \chi_{N-M}^{ [ \lambda_{L+1}  \dots \lambda_S ]}  { N \choose M } , \label{binorec2} 
\en
which can be easily proven starting from (\ref{xidef}).

\section{Extremizing the normalization ratio}
Bounds for $\chi_{N}$  and for $\chi_{N+1}/\chi_{N}$ are equivalent: Maximal (minimal) $M(m)$s maximize (minimize) the normalization factor $\chi_N$ as well as the ratio $\chi_{N+1}/\chi_N$, as we show in the following. 

Using Eq.~(\refnewton) in the main text, we can write $\chi_{N+1}=\chi_N+X_N$, where  $X_N \ge 0$ is a monotonically increasing function of all $M(m)$ with $2 \le m \le N$, such that we have
\eq 
\begin{split} 
\frac{\text{d}} {\text{d}M(m) } \log (\chi_{N+1} ) &\ge \frac{\text{d}} {\text{d}M(m) } \log (\chi_{N}) \Leftrightarrow \frac{1}{ \chi_{N+1}} \frac{\text{d}\chi_{N+1}}{\text{d}M(m) } \ge \frac{1}{ \chi_{N}} \frac{\text{d}\chi_{N}}{\text{d}M(m) } , \label{inequalitychiratio}
\end{split}
\en
which implies
\eq 
\begin{split} 
\frac{\text{d}}{\text{d}M(m)} \frac{ \chi_{N+1}}{\chi_N} & = \frac 1 {\chi_N} \frac{\text{d}\chi_{N+1}}{\text{d}M(m)}  - \frac{\chi_{N+1}}{\chi_N^2} \frac{\text{d}\chi_{N}}{\text{d}M(m)}   \\
&= \frac 1 {\chi_N} \left( \frac{\text{d}\chi_{N+1}}{\text{d}M(m)}- \frac{\chi_{N+1}}{\chi_N} \frac{\text{d}\chi_N}{\text{d}M(m)} 
 \right) \ge 0 .
\end{split}
\en
In other words, the monotonic dependence of $\chi_N$ on all $M(m)$ is inherited by $\chi_{N+1}/\chi_N$.

\section{Upper bound in the purity $P$ and in the largest Schmidt coefficient $\lambda_1$}
By defining operations on the distributions of Schmidt coefficients analogous to those introduced in Refs.~\cite{dMunford1977,dTichy2012CB}, we find that the distribution $\vec \Lambda_{\text{max}}(\lambda_1,P)$ that maximizes the power-sums $M(m \ge 3)$ under the constraints $\lambda_j \le \lambda_1$ and $\sum_{j} \lambda_j^2=P$ is constructed as follows: The largest possible Schmidt coefficient $\lambda_1$ is repeated as often as possible -- as allowed by normalization and by the constrained purity $P=M(2)$ -- namely $(L-1)$ times, with $L = \left\lceil P/\lambda_1^2 \right\rceil $. The $L$th coefficient $\lambda_L$ is then chosen to be as large as possible, while normalization is ensured by the remaining $S-L$ smaller coefficients, 
\eq 
\lambda_1=\lambda_2=\dots = \lambda_{L-1} \ge  \lambda_L \ge \lambda_{L+1} = \dots = \lambda_{S-1}= \lambda_S .  \\
\en
We therefore need to solve the quadratic equation
\eq \begin{split} 
(L-1) \lambda_1 + \lambda_L + (S-L) \lambda_S &=1 , \\
(L-1) \lambda_1^2 + \lambda_L^2 + (S-L) \lambda_S^2 &=P  , 
\end{split}
\en
for $\lambda_L$ and $\lambda_S$:
\eq 
\begin{split} 
\lambda_L&=  \frac{ 1-(L-1)\lambda_1+R}{S+1-L} , \\ 
\lambda_{S}&=  \frac{1-(L-1)\lambda_1}{S+1-L} -\frac{R }{(S-L)(S+1-L)} ,  
\end{split}
\en
where  
\eq 
R = \sqrt{( S-L) ( P (S+1- L)-1 + ( L-1) \lambda_1 (2 - \lambda_1 S))} ,
\en
and, in order to ensure $\lambda_S, \lambda_L \ge 0$, $S$ needs to be chosen large enough,
\eq 
S > \frac{(L-1)(P-2\lambda_1)+1}{P-(L-1)\lambda_1^2}.
\en
By applying Eq.~(\ref{binorec2}), we can write the normalization factor as follows,
\eq 
\chi_N^{\vec \Lambda_{\text{max}}(\lambda_1,P)}= \sum_{M=0}^N \sum_{K=0}^{N-M} \chi_{M}^{[\lambda_1 \dots \lambda_{L-1}]} \chi_{K}^{[\lambda_L]} \chi_{N-M-K}^{[\lambda_{L+1} \dots \lambda_{S}]} \frac{N!}{M!K!(N-M-K)!}.
\en
In the limit $S \rightarrow \infty$, 
\eq 
\lambda_L \rightarrow \sqrt{ P-(L-1) \lambda_1^2} ,
\en
and the Schmidt coefficients $\lambda_{L+1}\dots \lambda_S$ are infinitesimally small and do not contribute to power sums $M(m \ge 2)$. Using Eq.~(\ref{singlelam2}), we find the normalization factor associated to those coefficients:
\eq
\lim_{S\to\infty} \chi_{N-M-K}^{[\lambda_{L+1} \dots \lambda_{S}]} = \lim_{S\to\infty} \left(\frac{\lambda_{{\Sigma}}}{S-L}\right)^{N-M-K} \frac{(N-M-K+S-L-1)!}{(S-L-1)!} = \lambda_{{\Sigma}}^{N-M-K} \label{limitS}
\en
where
\eq
\lambda_{{\Sigma}}= 1- (L-1) \lambda_1- \lambda_L ,
\en
is the $S$-independent sum of all infinitesimal coefficients $\lambda_S$. Applying (\ref{singlelam2}) again, the normalization factor becomes
\eq 
\chi_N^{\vec \Lambda_{\text{max}}(\lambda_1,P)} = N! \sum_{M=0}^N \sum_{K=0}^{N-M} \lambda_1^{M} \lambda_L^{K} \lambda_{{\Sigma}}^{N-M-K} {M+L-2 \choose L-2} \frac{1}{(N-M-K)!}.
\en
The sum over $K$ may be recognized as the expansion of an incomplete $\Gamma$-function \cite{dOlver:2010:NHMF}, 
\eq
\begin{split}  
\chi_N^{\vec \Lambda_{\text{max}}(\lambda_1,P)}&= N! e^{\frac{\lambda_\Sigma}{\lambda_L}} \sum_{M=0}^N  \frac{\lambda_1^M \lambda_L^{N-M}}{(N-M)!} {M+L-2 \choose L-2} \Gamma\left(1+N-M,\frac{\lambda_\Sigma}{\lambda_L}\right).
\end{split}
\en

\section{Lower bound in the purity $P$ and in the largest Schmidt coefficient $\lambda_1$}
In analogy to the maximizing distribution, we find the distribution $\vec \Lambda_{\text{min}}(\lambda_1,P)$ that \emph{minimizes} the $M(m)$ by choosing as few coefficients $\lambda_2\dots \lambda_S$ as possible, namely $S=1 + \left \lceil \frac{(1-\lambda_1)^2}{P-\lambda_1^2} \right \rceil $, with 
\eq
 \lambda_1 \ge \lambda_2=\dots = \lambda_{S-1} \ge \lambda_S . 
\en
We therefore need to find $\lambda_2$ and $\lambda_S$ that fulfill  the quadratic equation
\eq 
\begin{split} 
 \lambda_1 + (S-2) \lambda_2  + \lambda_S &=1 , \\
\lambda_1^2 + (S-2) \lambda_2^2 +  \lambda_S^2 &=P   ,
\end{split}
\en
which is solved by 
\eq 
\begin{split} 
\lambda_{2, \dots, S-1}&=  \frac{1-\lambda_1}{S-1} + \frac{R^\prime}{(S-2)(S-1)}  , \\ 
\lambda_S&=   \frac{1-\lambda_1-R^\prime}{S-1} ,  \label{LowerBoundLambdaS} 
\end{split}
\en
where
\eq 
R^\prime= \sqrt{(S - 2) ( \lambda_1 (2 - S \lambda_1) + (S - 1) P - 1)} .
\en
By recourse to (\ref{singlelam2}) and (\ref{binorec2}), the normalization factor for $\vec \Lambda_{\text{min}}(\lambda_1,P)$ becomes
\eq
\chi_N^{\vec \Lambda_{\text{min}}(\lambda_1,P)} = N! \sum_{M=0}^N \sum_{K=0}^{N-M} \lambda_1^{K} \lambda_2^{M} \lambda_S^{N-M-K} {M+S-3 \choose S-3} , \label{ChiLowerBoundLP}
\en
where the sum over $K$ evaluates to 
\eq
\chi_N^{\vec \Lambda_{\text{min}}(\lambda_1,P)} = N! \sum_{M=0}^N \lambda_2^M \frac{\lambda_1^{1+N-M}-\lambda_S^{1+N-M}}{\lambda_1-\lambda_S} {M+S-3 \choose S-3}  . \label{ChiLowerBoundLPAnalitEva}
\en

\section{Bounds in the purity $P$}
\subsection{Lower bound}
The distribution that minimizes the $M(m)$ under the constraint $M(2)=P$ is obtained for $\lambda_1= \lambda_{1,\text{min}}(P)$, which yields the \emph{uniform} distribution introduced in Ref.~\cite{dTichy2012CB}. 
We then have 
\eq
\lambda_{1,\text{min}}(P)=\lambda_{1}=\lambda_{2}= \dots = \lambda_{S-1}, ~ ~ ~ \lambda_{S}= {1-\lambda_1 (S-1)}=\frac{1 - \sqrt{(S-1)(S P -1)}}{S}, 
 ~\text{with}~ S=\left\lceil \frac{1}{P} \right\rceil ,
\en
such that the normalization factor becomes 
\eq 
\chi_N^{\vec \Lambda_{\text{min}}(\lambda_{1,\text{min}}(P),P)} = \frac{ N!}{(S-2)!}\sum_{M=0}^N \lambda_1^{M} \lambda_S^{N-M} \frac{(M+S-2)!}{M! } \stackrel{(a)}{ \ge} P^N \frac{\Gamma\left(N+\frac{1}{P}\right)}{\Gamma\left(\frac{1}{P}\right)} , \label{lowerbound}
\en
where the lower bound $(a)$ is obtained by ignoring all but the last summand in the sum over $M$. 
The  numerical evaluation of (\ref{lowerbound}) for the uniform distribution is most conveniently done via 
\eq
\chi_N^{\vec \Lambda_{\text{min}}(\lambda_{1,\text{min}}(P),P)}= \lambda_{1}^{N}  \frac{(N+S-2)!}{(S-2)!}~ {}_2F_1\left(1, -N, 2 - N - S, \frac{\lambda_{S}}{\lambda_{1}}\right),
\en
where ${}_2F_1$ is the ordinary hypergeometric function \cite{dOlver:2010:NHMF}.

\subsection{Upper bound}
The power sums $M(m)$ are maximized by $\lambda_{1}\rightarrow \lambda_{1,\text{max}}(P)$. We obtain the \emph{peaked} distribution with $S$ non-vanishing coefficients \cite{dTichy2012CB}
\eq
\lambda_{1,\text{peak}} = \frac{1 + \sqrt{(S-1)(S P -1)}}{S}, ~ ~ \lambda_{j \in \{ 2\dots S\},\text{peak}} = \frac{1-\lambda_{1,\text{peak}}}{S-1} . 
\en
In the  limit $S\to\infty$, the only non-vanishing coefficient is $\lambda_{1,\text{peak}}=\lambda_{1,\text{max}} = \sqrt{P}$, the sum of the other coefficients converges to $1-\sqrt{P}$. Adapting (\ref{limitS}), the normalization factor becomes 
\eq
\chi_N^{\vec \Lambda_{\text{max}}(\lambda_{1,\text{max}}(P),P)} = N! \sum_{M=0}^N P^{M/2} (1-\sqrt{P})^{N-M} \frac{1}{(N-M)!}  ,
\en
which can be expressed as an incomplete $\Gamma$-function \cite{dOlver:2010:NHMF}, \eq
\chi_N^{\vec \Lambda_{\text{max}}(\lambda_{1,\text{max}}(P),P)} = e^{\frac{1}{\sqrt{P}}-1} P^{\frac N 2} \Gamma\left(1+N,\frac{1}{\sqrt{P}}-1\right) .
\en
Using $\Gamma(x,y)= (x-1) \Gamma(x-1,y) + y^{x-1} e^{-y}$, one finds a simpler upper bound for the normalization ratio:
\eq
\begin{split} 
\frac
{\chi_{N+1}^{\vec \Lambda_{\text{max}}(\lambda_{1,\text{max}}(P),P)}}
{\chi_N^{\vec \Lambda_{\text{max}}(\lambda_{1,\text{max}}(P),P)} } 
&\leq \sqrt{P} \frac{\Gamma(2+N,\frac 1{\sqrt{P}}-1)}{\Gamma(1+N,\frac 1{\sqrt{P}}-1)} = \sqrt P \frac{(1+N)\Gamma \left(1+N,\frac 1{\sqrt{P}}-1 \right)+\left(\frac 1 {\sqrt{P}}-1 \right)^{N+1} e^{-\frac 1{\sqrt{P}}+1}}{\Gamma \left(1+N,\frac 1{\sqrt{P}}-1 \right)}  \\
& = \sqrt{P} (N + 1) + \frac{\sqrt{P}^{N+1} \left(\frac 1 {\sqrt{P}}-1 \right)^{N+1} } { \chi_N^{\vec \Lambda_{\text{max}}(\lambda_{1,\text{max}}(P),P)} } \le  \sqrt P N + 1 ,
\end{split}
\en
which is recovered by the normalization factor
\eq 
\chi_N^{\vec \Lambda_{\text{max}}(\lambda_{1,\text{max}}(P),P)} 
\le P^\frac{N}{2}  \frac{\Gamma\left(N+\frac{1}{\sqrt{P}}\right)}{\Gamma\left(\frac{1}{\sqrt{P}}\right)} .
\en

\section{Bounds in the largest Schmidt coefficient $\lambda_1$}

\subsection{Lower bound} 
Given $\lambda_1$, the power sums $M(m)$ are minimized for $P_\text{min}(\lambda_1)=\lambda_1^2$ in the limit $S\to\infty$. In this limit, the coefficient $\lambda_S$ in (\ref{LowerBoundLambdaS}) vanishes and the sum of the coefficients $\lambda_2,\dots,\lambda_{S-1}$ remains finite and equal to $1-\lambda_1$. The 
resulting  normalization ratio can be computed using (\ref{ChiLowerBoundLP}),
\eq
\chi_N^{\vec \Lambda_{\text{min}}(\lambda_1,P_{\text{min}}(\lambda_1))} = N! \sum_{M=0}^N \lambda_1^{M} (1-\lambda_1)^{N-M} \frac{1}{(N-M)!} \stackrel{(b)} {\ge} \lambda_1^N N! ~,
\en
where the inequality $(b)$ is obtained by only allowing for the last summand in the sum over $M$. An efficient numerical evaluation is possible by using the representation as an incomplete $\Gamma$-function, 
\eq
\chi_N^{\vec \Lambda_{\text{min}}(\lambda_1,P_{\text{min}}(\lambda_1))} = e^{\frac{1}{\lambda_1}-1} \lambda_1^N \Gamma\left(1 + N, \frac{1}{\lambda_1}-1\right) .
\en

\subsection{Upper bound} 
Power sums $M(m)$ are maximized for $P=P_{\text{max}}(\lambda_1)$. 
Such a distribution is constructed by choosing the highest multiplicity possible  of $\lambda_1$, i.e. there are $\left\lfloor \frac{1}{\lambda_1} \right\rfloor$ coefficients of magnitude $\lambda_1$, and one of magnitude $1-\left\lfloor \frac{1}{\lambda_1} \right\rfloor \lambda_1$. The resulting normalization factor reads 
\eq
\chi_N^{\vec \Lambda_{\text{max}}(\lambda_1,P_\text{max}(\lambda_1))} = N! \sum_{M=0}^N \left(1-\left\lfloor\frac{1}{\lambda_1}\right\rfloor \lambda_1\right)^{M} \lambda_1^{N-M} 
{ N-M+\left\lfloor\frac{1}{\lambda_1}\right\rfloor-1 
\choose N-M } ,
\en
whose  numerical evaluation  is most efficient using
\eq
\chi_N^{\vec \Lambda_{\text{max}}(\lambda_1,P_\text{max}(\lambda_1))} = \lambda_1^N \frac{ \left(N+ \left\lfloor \frac{1}{\lambda_1} \right\rfloor-1 \right)!}{ \left(\left\lfloor \frac{1}{\lambda_1} \right\rfloor-1 \right)!} ~ {}_2F_1\left(1,-N;1-N-\left\lfloor \frac{1}{\lambda_1} \right\rfloor ;\frac{1-\lambda_1 \left\lfloor \frac{1}{\lambda_1} \right\rfloor}{\lambda_1} \right) .
\en

When $\lambda_1$ is in the range between the inverses of two consecutive integers, $1/L < \lambda_1 < 1/(L-1)$, we can bind the normalization ratio by the one obtained for these extremes, $\lambda_1=1/L$ and $\lambda_1=1/(L-1)$:
\eq
\frac{N}{L} + 1<
\frac{\chi_{N+1}^{\vec \Lambda_{\text{max}}(\lambda_1,P_\text{max}(\lambda_1))}}{\chi_N^{\vec \Lambda_{\text{max}}(\lambda_1,P_\text{max}(\lambda_1))}}< \frac{N}{L-1} + 1 . \label{upperlower}
\en
In that range, the normalization ratio is convex in $\lambda_1$, and the linear interpolation between the upper and lower bounds in (\ref{upperlower}) yields a simpler upper bound:
\eq
 \frac{\chi_{N+1}^{\vec \Lambda_{\text{max}}(\lambda_1,P_\text{max}(\lambda_1))}}{\chi_N^{\vec \Lambda_{\text{max}}(\lambda_1,P_\text{max}(\lambda_1))}} \le N \lambda_1 + 1 .
 \en
This relation is also inherited by the normalization factor, and 
\eq
\chi_{N}^{\vec \Lambda_{\text{max}}(\lambda_1,P_\text{max}(\lambda_1))}
\le 
\prod_{j=0}^{N-1} \left( j \lambda_1 + 1 \right)  = 
\lambda_1^N \frac{\Gamma\left(N+\frac{1}{\lambda_1}  \right)}{\Gamma\left(  \frac{1}{\lambda_1}  \right)},
\en

\section{Overview over all  bounds}
In order to summarize the established results, we reproduce the bounds for the normalization factor $\chi_N^{\vec \Lambda}$ of a distribution $\vec \Lambda$ with $M(2)=P$ and with a maximal Schmidt coefficient $\lambda_1$:
\eq
\left.
\begin{array}{r} 
P^N \frac{\Gamma\left(N+\frac{1}{P}\right)}{\Gamma\left(\frac{1}{P}\right)} \stackrel{(c)}{\le} \chi_N^{\vec \Lambda_{\text{min}}(\lambda_{1,{\text{min}}}(P),P)}  \\
\lambda_1^N N! \stackrel{(d)}{\le} \chi_N^{\vec \Lambda_{\text{min}}(\lambda_1, P_{\text{min}}(\lambda_1) )} 
\end{array} \right\}
\le \chi_N^{\vec \Lambda_{\text{min}}(\lambda_1,P)} \le \chi_N \le \chi_N^{\vec \Lambda_{\text{max}}(\lambda_1,P)} 
\le
\left\{ \begin{array}{l} 
\chi_N^{\vec \Lambda_{\text{max}}(\lambda_{1,{\text{max}}}(P),P)} \stackrel{(d)}{\le} P^\frac{N}{2}  \frac{\Gamma\left(N+\frac{1}{\sqrt{P}}\right)}{\Gamma\left(\frac{1}{\sqrt{P}}\right)} \\
\chi_N^{\vec \Lambda_{\text{max}}(\lambda_1,P_{\text{max}}(\lambda_1))} \stackrel{(c)}{\le} \lambda_1^N \frac{\Gamma\left(N+\frac{1}{\lambda_1}\right)}{\Gamma\left( \frac{1}{\lambda_1}\right)}
\end{array} \right. ,
\en
The visual arrangement of the bounds reflects the graphical representation given in Fig.~2 in the main text. Inequalities marked with $(c)$ are efficient, while those marked with $(d)$ become tight only for $N \gg 1/\lambda_1$ or $N \gg 1/\sqrt P$, respectively.

\end{widetext}

\end{document}